\documentclass[fleqn,usenatbib]{mnras}

\usepackage{newtxtext,newtxmath}

\usepackage[T1]{fontenc}

\DeclareRobustCommand{\VAN}[3]{#2}
\let\VANthebibliography\thebibliography
\def\thebibliography{\DeclareRobustCommand{\VAN}[3]{##3}\VANthebibliography}

\usepackage{graphicx}	
\usepackage{amsmath}	
\usepackage{booktabs, caption}
\usepackage[flushleft]{threeparttable}

\newcommand{\rc}{\color{black}}

\title[Gas tori formed by planetary mass loss]{Observational Signatures of Circumstellar Gas Tori Formed by Planetary Mass-Loss from Close-In Exoplanets}

\author[Ethan Schreyer]{
Ethan Schreyer,$^{1,2}$\thanks{E-mail: eschreye@ucsc.edu}
Ruth Murray-Clay$^{2}$
\\
$^{1}$Astrophysics Group, Imperial College London, Blackett Laboratory, Prince Consort Road, London SW7 2AZ, UK\\
$^{2}$Department of Astronomy and Astrophysics, University of California, Santa Cruz, CA 95064, USA
}

\date{Accepted XXX. Received YYY; in original form ZZZ}

\pubyear{2025}

\begin{document}
\label{firstpage}
\pagerange{\pageref{firstpage}--\pageref{lastpage}}
\maketitle

\begin{abstract}
Close-in exoplanets with H/He atmospheres often undergo hydrodynamic escape. In extreme cases, it is hypothesized that the mass loss can be high enough for the escaping planetary material to wrap around the star, forming a long-lasting circumstellar torus. In this work, we develop a physical model of such circumstellar tori and use a ray tracing scheme to calculate the attenuation of stellar light passing through them. We show that the presence of a circumstellar torus  significantly increases the equivalent width of the observed stellar He\,\textsc{i} 10830~\AA\ line. When combined with observations of the star's Ca\,\textsc{ii} H \& K lines, these systems can typically be distinguished from field stars. Based on these results, we propose a survey of stars hosting close-in planets, combining observations of the He\,\textsc{i} 10830~\AA\ and Ca\,\textsc{ii} H \& K lines to search for circumstellar tori generated from planetary mass-loss in these systems. 
\end{abstract}




\section{Introduction}

Close-in exoplanets with extended hydrogen and helium envelopes often undergo significant mass loss. Mass loss can be directly observed by tracing the escaping gas using transit spectroscopy. Considerable observational effort has been dedicated to identifying and studying mass loss across a range of exoplanetary systems, primarily using Ly $\alpha$ or He\,\textsc{i} 10830 \AA~transits \citep[e.g.,][]{Vidal-Madjar2003, Lecavelier-des-Etangs2004, Eherenreich2015, Bourrier2017, Ben-Jaffel2022, Spake2018, Allart2018, Nortmann2018, ZhangM2023}. These observations have revealed a diverse set of transit signatures that often exhibit spatial and kinematic asymmetry \citep[e.g.,][]{Eherenreich2015, Lavie2017, Bourrier2018, Spake2021}, indicating that there are significant inter-system variations in the spatial and velocity structure of the escaping gas. These variations are thought to result from the broad range of parameters that shape the structure of the outflowing gas, including the stellar wind, planetary and stellar magnetic fields, radiation pressure, and the intrinsic properties of the planet \citep[e.g.,][]{Matsakos2015, McCann2019, Khodachenko2019, Carolan2021b}.

Whilst some observations predominantly trace absorption from gas within the planet's Hill sphere, others detect absorption occurring many hours outside the optical transit, revealing that the escaping gas extends far outside the Hill sphere of the planet \citep[e.g.,][]{Eherenreich2015, Bourrier2018, Zhang2023, Gully-Santiago2023}. For typical planetary mass-loss rates, the escaping gas is expected to be rapidly dispersed by the stellar wind or accreted onto the star, resulting in the periodic transit signatures observed. However, the detection of such extended gas structures raises the possibility that, under certain conditions, the gas is not dispersed and instead accumulates, {\rc over many orbital timescales}, into a {\rc long-lived} circumstellar torus. Although this is expected to be a rare scenario, it may occur for extremely close-in planets undergoing extreme photoevaporative mass loss or Roche lobe overflow \citep[e.g.,][]{Valsecchi2014, Valsecchi2015, Koskinen2022}.  

{\rc One way to test for the presence of circumstellar gas around stars hosting close-in planets is to search for absorption lines from this gas imprinted on the stellar spectrum. The circumstellar gas is inherited from the gas that escapes the planet and, therefore, is hot ($\sim 10^4$ K) and predominantly atomic/ionic \citep[e.g.,][]{Yelle2004}. Consequently, many of the strong absorption lines, such as hydrogen Ly $\alpha$ or metal resonance lines, from the circumstellar gas will be in the ultraviolet (UV). Unfortunately, strong interstellar extinction of Ly $\alpha$ completely obscures the stellar Ly $\alpha$ for stars beyond 100 pc \citep{Wood2005a}, limiting the observable sample. This is a particular challenge given that circumstellar tori produced by planetary mass-loss are expected to be intrinsically rare. 

Optical and near-infrared (near-IR) lines can also be used to search for absorption lines from the circumstellar gas. Balmer lines, the He\,\textsc{i} 10830 \AA~line and prominent optical resonance lines, such as the Ca\,\textsc{ii} H \& K lines are all strong candidates for detecting this gas. A key advantage of optical and near-IR lines is that they are accessible from the ground, enabling high-resolution spectroscopy and allowing observations of a much larger sample of stars than possible in the UV. 

The primary challenge in unambiguously detecting a circumstellar torus lies in distinguishing its absorption features from intrinsic stellar lines. The strong absorption lines probed by the circumstellar gas are also typically stellar chromospheric lines, and are often polluted by interstellar medium extinction. Since the circumstellar gas is expected to be on approximately circular orbits and the radial velocity of gas transiting the star is low (otherwise the gas would escape the system), the circumstellar absorption lines and stellar lines are co-located in wavelength. This means that there is typically no accurate template for the spectral region where circumstellar gas absorbs. Consequently, it is generally not possible to disentangle the stellar lines from the circumstellar absorption. There may be cases where circumstellar absorption is very broad or deep, or where an exceptionally good comparison template allows it to be clearly identified; however, such situations are not typical. Instead, to search for circumstellar gas, one can compare a star's spectra to a reference population and look for anomalies. This situation contrasts with observations of circumstellar gas from supergiant mass loss, where, for example, Ca\,\textsc{ii} H \& K absorption is more easily separated from stellar lines \citep[e.g.,][]{Reimers1977} due to it's high radial velocity, or where the significantly higher columns densities mean that elements such as Fe\,\textsc{i}, Ti\,\textsc{ii}, or Sr\,\textsc{ii} produce optical absorption lines that do not coincide with chromospheric stellar lines, and are distinguishable from the stellar photospheric contribution \citep[e.g.,][]{Weymann1962}.} 

An example of a system with unusual spectra is WASP-12. The central star is orbited by an ultra-hot Jupiter, WASP-12 b, on a tight 1.09-day orbit. Near-ultraviolet (NUV) transit spectroscopy of WASP-12 b revealed a deeper transit depth compared to the optical, attributed to absorption by a range of escaping metals: most prominently Mg\,\textsc{ii} and Fe\,\textsc{ii} \citep{Fossati2010, Haswell2012}. Curiously, it was observed that WASP-12 lacks emission in the cores of the resonant Mg\,\textsc{ii} lines, regardless of the orbital phase of WASP-12 b. Stellar Mg\,\textsc{ii} lines typically appear as broad photospheric absorption features with narrow core emission originating from the chromosphere. Such chromospheric core emission is prominent in stars of similar age and rotational velocity, {\rc for example HD 102634 and HD 107213 \citep[][]{Haswell2012}}, making its absence in WASP-12 particularly unusual. Subsequent optical observations revealed that the same is true for the resonant Ca\,\textsc{ii} lines \citep[][]{Fossati2013}, which are formed similarly. One possible explanation for these anomalies is that the gas escaping from WASP-12b has accumulated in a circumstellar torus, and thus attenuates the cores of both the Mg\,\textsc{ii} and Ca\,\textsc{ii} lines. 

Inspired by the observations of WASP-12, several studies have sought to identify similar systems and assess how common the circumstellar gas fed by a planet undergoing atmospheric escape might be \citep[e.g.,][]{Staab2017, DohertyThesis}. These investigations conducted spectroscopic observations of the stellar Ca\,\textsc{ii} H \& K lines, from which they calculated the star’s log\,$R_{\text{HK}}'$. This quantity measures the flux in the cores of the Ca\,\textsc{ii} H \& K lines, which originates in the chromosphere, relative to the star’s bolometric flux. As such, log\,$R_{\text{HK}}'$ is typically used as an indicator of stellar chromospheric activity. It has been proposed that, as in the case of WASP-12 b, an anomalously low log\,$R_{\text{HK}}'$ may signal the presence of circumstellar gas absorbing in the Ca\,\textsc{ii} H \& K line cores \citep[e.g.,][]{Staab2017, Haswell2020}. The OU-SALT survey \citep{DohertyThesis} found that approximately one-third of stars hosting close-in transiting planets exhibited a log\,$R_{\text{HK}}'$ lower than that of the Sun (log\,$R_{\text{HK}}' < -5.1$), compared to just 2\% of field stars, providing support for this hypothesis. A link between stellar log\,$R_{\text{HK}}'$ and planetary surface gravity had already been noted by \citet{Hartman2010} and \citet{Figueira2014}, who found a positive correlation between the surface gravity of hot Jupiters and the log\,$R_{\text{HK}}'$ of these planets host stars. \citet{Lanza2014} suggested that this correlation may arise because low-gravity planets undergo more extreme mass loss, leading to increased circumstellar Ca\,\textsc{ii} densities and consequently lower observed log\,$R_{\text{HK}}'$ values. 

Together, these studies suggest an astrophysical link between a star’s log\,$R_{\text{HK}}'$ and the presence of a close-in planet. However, it remains unclear whether these observations are driven by planetary mass-loss and the subsequent formation of circumstellar gas, intrinsic differences between close-in planet hosts and typical main-sequence stars, or other, as-yet unidentified physical mechanisms.

{\rc In this work, we explore methods to test whether the observed link between a star’s log\,$R_{\text{HK}}'$ and the presence of a close-in planet arises due to the attenuation of light by a circumstellar torus, generated by planetary atmospheric mass loss. Our main finding is that the He\,\textsc{i} 10830 \AA~line is an excellent diagnostic of circumstellar material, and we therefore propose its use as an observational tracer alongside the Ca\,\textsc{ii} H \& K lines. To demonstrate this, we first develop a simple analytical model of the circumstellar gas along with a ray-tracing scheme that calculates the optical depth of the circumstellar gas as a function of wavelength. We then calculate how a circumstellar torus would modify the observed  He\,\textsc{i} 10830 \AA~line and Ca\,\textsc{ii} H \& K lines for an ensemble of stars. We find that stars hosting circumstellar tori are spatially separated from a reference population without circumstellar tori in He\,\textsc{i} 10830~\AA~EW-log\,$R_{\text{HK}}'$ space, indicating that the combination of these two tracers provides a strong observational signature of circumstellar material generated by planetary mass loss. Finally, we find that circumstellar material affects the observed planetary transit in the He\,\textsc{i} 10830 \AA~line, reducing the meaured transit depth and potentially eliminating the transit signal altogether.}

\section{Observational Background}

\begin{figure}
\centering
\includegraphics[width=0.47\textwidth]{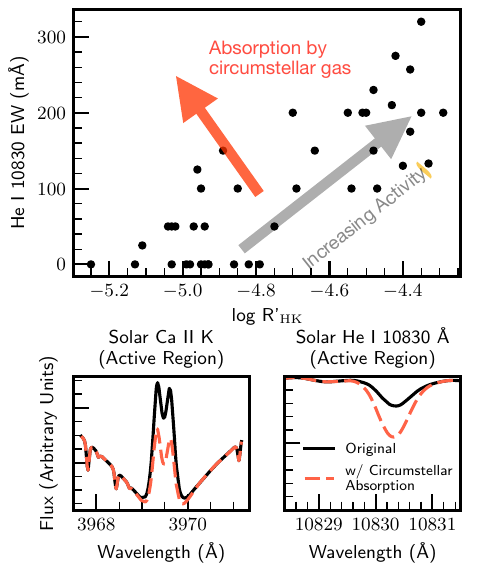}
\caption[Helium, Calcium Correlation]{Top Panel: Equivalent width of the He\,\textsc{i} 10830 \AA~feature versus the chromospheric Ca\,\textsc{ii} emission index log\,$R_{\text{HK}}'$, for a sample of dwarf FGK stars with (B-V) $> 0.47$, compiled by \citet{Smith2016}. A strong correlation is observed, indicated by the grey arrow, which shows the direction of increasing stellar activity. The red arrow illustrates how circumstellar gas absorption would shift a star's observed properties away from the stellar activity trend. Bottom Left Panel: Example spectrum of the core of the solar Ca\,\textsc{ii} K line in an active region (black solid line), from \citet{White1981}. The chromospheric emission appears within a broad photospheric absorption feature. Absorption by circumstellar gas would reduce the observed core emission, as illustrated by the red dashed line, and hence the measured log\,$R_{\text{HK}}'$. Bottom Right Panel: Same as the left panel, but for the solar He\,\textsc{i} 10830 \AA~line in an active region from \citet{Sanz-Forcada2008}. Absorption from circumstellar gas would deepen the line, and hence increase the measured equivalent width.}
\label{fig:he_ca_correlation}
\end{figure}

{\rc To identify absorption from circumstellar gas, it is first necessary to understand the intrinsic stellar spectrum in the wavelength regions where the gas absorbs. Stellar surveys provide valuable information on the strength and shape of stellar spectral lines, and how these features correlate with other stellar properties. These surveys have focused on a broad range of wavelengths ranging from the ultraviolet \citep[e.g.,][]{Ayres1981, France2016}, through the optical \citep[e.g.,][]{Wilson1968, Boro-Saikia2018} to the near-IR \citep[e.g.,][]{Zarro1986, Sanz-Forcada2008}. Since ultraviolet observations can only be obtained from space, optical and near-IR surveys are for more extensive than their ultraviolet counterparts, and therefore optical and near-IR lines are typically better characterized across a wide range of stars, making it easier to identify the imprint of circumstellar absorption using these lines. For this reason, we focus on the  Ca\,\textsc{ii} H \& K lines (optical) and He\,\textsc{i} 10830 \AA~line (near-IR), both of which are potential tracers of circumstellar gas and have been the subject of extensive surveys \citep[e.g.,][]{Wilson1968, Zirin1982,Zarro1986,Obrien1986, Sanz-Forcada2008, Smith2016, Boro-Saikia2018}. 

The Ca\,\textsc{ii} H \& K resonance lines appear as broad photospheric lines with narrow emission cores formed in the upper chromosphere. The strength of the emission cores are a measure of the activity of the star. Because a star's activity level cannot be predicted solely from its spectral type, the strengths of the Ca\,\textsc{ii} H \& K emission cores can vary significantly among otherwise similar stars. As a result, constructing a reliable template for the Ca\,\textsc{ii} H \& K profiles requires knowledge of additional stellar properties (e.g., age, rotational period). Additionally observations of the Ca\,\textsc{ii} H \& K lines also suffer from extrinsic absorption in the intervening ISM. Therefore, pulling out any circumstellar absorption component is typically uncertain. Despite this, the Ca\,\textsc{ii} H \& K lines have been proposed as an indicator of circumstellar gas \citep[e.g.,][]{Haswell2012, Staab2017}. Stars are observed to exhibit a minimum, or basal, level of chromospheric emission in the Ca\,\textsc{ii} H \& K lines \citep[][]{Perez-Martinez2014}, analogous to that seen during the quiet Sun \citep{Schroder2012}. \citet{Staab2017, Haswell2020} suggest that some stars exhibiting Ca\,\textsc{ii} H \& K chromospheric emission below this basal limit (log\,$R_{\text{HK}}' \sim -5.1$) are shrouded by circumstellar material, formed from mass loss from a close-in planet. 

The He\,\textsc{i} 10830 \AA~line is a chromospheric/transition region (typically) absorption line\footnote{It can be an emission line for very young or active stars} produced by transitions between the between the metastable $2^3$s state and $2^3$p state. For inactive stars, $L_\text{X}/L_{\text{bol}} \lesssim 10^{-4}$, the equivalent width of the helium line correlates with the X-ray flux \citep{Zarro1986, Sanz-Forcada2008, Smith2016}. Therefore, like the Ca\,\textsc{ii} H \& K lines, the equivalent width of the He\,\textsc{i} 10830 \AA~line varies significantly depending on the activity of the star. This makes constructing a reliable template of the He\,\textsc{i} 10830 \AA~line for a specific star difficult. 

Interpreting the circumstellar contribution to the observed Ca\,\textsc{ii} H \& K or He\,\textsc{i} 10830~\AA~lines is challenging when these lines are analysed individually. First, diagnosing the presence of circumstellar gas is difficult when the absorption is too weak to alter the line profile sufficiently to stand out as an outlier. Additionally, other factors can make the line profile appear unusual. For example, if the observed Ca\,\textsc{ii} H \& K emission is low, it is not clear whether the star is intrinsically unusually inactive or the star hosts a circumstellar torus. To help circumvent these problems, we can exploit the fact that the strength of the stellar He\,\textsc{i} 10830~\AA~absorption line is highly correlated with strength of chromospheric emission in Ca\,\textsc{ii} H \& K lines (log\,$R_{\text{HK}}'$), as shown in Figure~\ref{fig:he_ca_correlation} \citep{Smith2016}. If a star exhibits an anomalously low chromospheric Ca\,\textsc{ii} H \& K emission that is driven by intrinsic stellar processes, such as activity, then the He\,\textsc{i} 10830~\AA~absorption line and Ca\,\textsc{ii} H \& K chromospheric emission should still follow this line, as we would expect the He\,\textsc{i} 10830~\AA line to be equally affected. In contrast, if the anomalous behaviour arises from something extrinsic to the star, such as a circumstellar torus, then the stars would likely deviate from this line. 

A sketch of this idea is shown in Figure~\ref{fig:he_ca_correlation}. In the top panel of Figure~\ref{fig:he_ca_correlation}, we have plotted the He\,\textsc{i} 10830~\AA~EW and log\,$R_{\text{HK}}'$ of sample of dwarf FGK stars with (B-V) $>$ 0.47, compiled by \citet{Smith2016}. In the bottom panel, we plot example Solar Ca\,\textsc{ii} K (left, \citealt{White1981}) and He\,\textsc{i} 10830 \AA~(right, \citealt{Sanz-Forcada2008}) lines in active regions. To illustrate how absorption by a circumstellar torus might modify the observed profiles, we show these lines attenuated by a Gaussian profile with standard deviation 0.25 \AA (dashed red line). For the Ca\,\textsc{ii} K line the Gaussian has normalized amplitude of 0.3 and is centred at the 3969.59 \AA~, the rest wavelength of this transition. For the He\,\textsc{i} 10830 \AA~line, we use two Gaussians. The first has a normalized amplitude of 0.3 and is centred at 10833.27 \AA~, the central wavelength of the two blended triplet transitions  with rest wavelengths of 10833.31 and 10833.22 \AA~. The second has a normalized amplitude of 0.04 and is centred at 10832.06 \AA~, the rest wavelength of the other triplet transition. The amplitude ratio between the Gaussians is $\sim 8$, which reflects the relative strengths of these transitions and is consistent with the amplitude ratio for optically thin absorption.

The circumstellar torus attenuates the light from the star decreasing the observed Ca \,\textsc{ii} H \& K chromospheric emission (and hence log\,$R_{\text{HK}}'$) and increasing the depth of the He\,\textsc{i} 10830~\AA~absorption line. These changes move the star in He\,\textsc{i} 10830~\AA~EW-log\,$R_{\text{HK}}'$ space as shown by the red arrow in the the top panel in Figure~\ref{fig:he_ca_correlation}. The star moves in a direction orthogonal to the activity-driven correlation between He\,\textsc{i}10830~\AA~EW and log\,$R_{\text{HK}}'$. Therefore, even moderate attenuation would cause a star hosting a circumstellar torus to be an outlier.}

\section{Model of the circumstellar gas}\label{sec:model}

In this section, we develop a simple analytic model for the circumstellar gas produced by planets undergoing extreme mass loss. If the escaping gas is not dispersed by the stellar wind or radiation pressure, or rapidly accreted onto the star, 3D hydrodynamic simulations show that the gas initially forms a circumstellar torus \citep[e.g.,][]{Debrecht2018, McCann2019, Macleod2022}. Due to their high computational cost, these simulations were run for, at most, tens of planetary orbital periods, so the long-term evolution of the gas is uncertain. Viscous forces, magnetic interactions with the star, and gravitational interactions with the planet may subsequently alter the structure and distribution of the gas. However, we broadly expect the density of the circumstellar gas to increase with time. 

The accumulation of gas cannot proceed indefinitely---eventually sufficient gas should be present to shield the source planet from radiation that drives atmospheric escape, cutting off additional gas supply. We estimate that this occurs when the optical depth of gas between the star and planet to EUV ionizing radiation is unity, so that photoevaporation no longer operates. In this paper, we will assume that the density of the circumstellar gas is set by this limit, unless otherwise stated. However, we note that other mechanisms may be responsible for setting the limiting density of circumstellar gas. It is feasible that the density of the gas may exceed this EUV limit if planetary mass loss continues via X-ray driven mass loss or Roche lobe overflow. Another plausible density scale for the gas is accretion onto the star balancing planetary mass loss. Alternatively, it is possible that stellar coronal mass ejections are strong enough to partially or fully disperse the circumstellar gas. These could lead to variable gas structure, or if frequent enough may prevent the formation of this structure altogether. 

To construct a simple model of the circumstellar gas, we make the following assumptions regarding its geometry and composition:\\

\noindent (i) {\rc The circumstellar gas is arranged in a toroidal ring centred on the star that has a radius equal to the planet’s semi-major axis. The toroidal ring has a fixed vertical height and depth.}\\

\noindent (ii) The gas within the torus has a uniform density.\\

\noindent (iii) The gas is optically thin to EUV radiation.\\

\noindent (iv) The gas within the torus is in photoionization-recombination equilibrium.\\

\begin{figure*}
\centering
\includegraphics[width=0.95\textwidth]{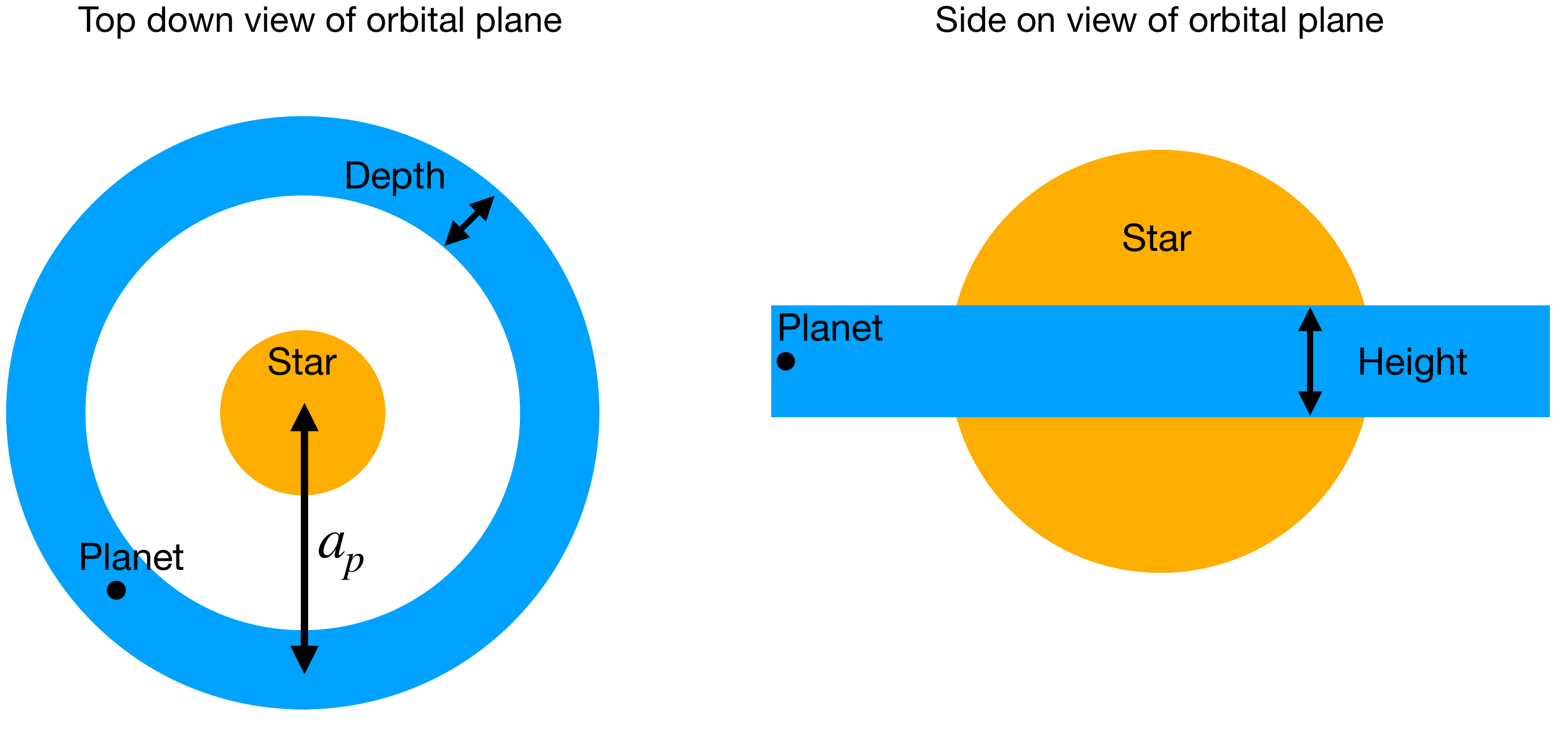}
\caption[Circumstellar Torus Schematic]{{\rc A schematic diagram of the circumstellar gas torus, that has formed due to the accumulation of escaping gas from the planet over many orbits. The left and right panel shows the system geometry when observed top-down (left) and side on (right).}}
\label{fig:schematic}
\end{figure*}

\noindent {\rc A schematic diagram of this simple model is shown in Figure~\ref{fig:schematic}.} We acknowledge that the structure of the circumstellar gas will be more complex than is assumed in this simple model. However, we expect that the qualitative features of the circumstellar gas should be preserved.  We aim to use this simple formulation to characterize the first-order observational signatures of accumulating gaseous tori, so that we may construct observational tests to prove or disprove their existence, deduce their basic properties, and, if discovered, motivate more detailed models.

\subsection{Formation of circumstellar gas}

Here, we estimate the parameter space where a circumstellar gas torus is expected to form. For a planet undergoing hydrodynamic escape, the velocity of the escaping gas, is comparable to its thermal velocity, and is much smaller than the planet’s orbital velocity. As a result, the gas retains nearly the same angular momentum and energy as the planet and is put into a nearly circular orbit around the star \citep{Owen2023, Schreyer2024b}. In the planet's frame of reference, the gas forms a tube-like structure leading and trailing the planet along its orbit, growing due to the relative velocity between it and the planet. Undisrupted, the leading and trailing legs of the gas grow so that the gas eventually forms a circumstellar torus \citep[e.g.,][]{Debrecht2018, McCann2019, Macleod2022}. However, the stellar wind, radiation pressure and stellar magnetic field can disrupt the gas, stopping this from occurring. For example, when the stellar wind is strong, the gas is shaped into a cometary tail trailing the planet, moving radially outwards. This is observed in the Lyman-$\alpha$ transits of GJ436 b \citep[e.g.,][]{Eherenreich2015, Lavie2017} and GJ3470 b \citep[e.g.,][]{Bourrier2018}. In the following sections, we estimate the conditions required for a complete gaseous circumstellar torus to develop and persist.   

\subsubsection{Stellar wind}

The ram pressure of the stellar wind imparts a force onto the escaped planetary gas. This force can disrupt the planetary gas in two ways, depending on the direction of incidence. It accelerates the planetary gas radially outwards, eventually causing it to fragment and be removed from the system \citep[e.g.][]{Matsakos2015, McCann2019, Khodachenko2019}. The stellar wind can also remove angular momentum from the planetary gas, so that it is eventually accreted onto the star \citep[e.g.,][]{Matsakos2015}. For simplicity, we estimate the conditions required to form a circumstellar torus, considering a radial stellar wind.

Following \citet{Owen2023, Schreyer2024b}, in a frame co-rotating with the planet, the force per unit length on the tube of escaped planetary gas by the stellar wind is:
\begin{align}
\boldsymbol{f_{sw}} = 2H\rho_*(\boldsymbol{u}_{*}  - \boldsymbol{u} \boldsymbol{\cdot} \hat{\boldsymbol{u}}_{*})^2 \hat{\boldsymbol{u}}_{*}
\end{align}
where $\rho_*$ is the density of the stellar wind, $\boldsymbol{u_*}$ is the velocity of the stellar wind, $\boldsymbol{u}$ is velocity of the planetary gas and $H$ is the height of the tube. The mass of the tube per unit length is $\approx \frac{\dot{M}_p}{u}$, therefore the acceleration of the planetary gas by the stellar wind is:
\begin{align}
\boldsymbol{a_\text{sw}} = \frac{2Hu\rho_*(\boldsymbol{u}_* - \boldsymbol{u} \boldsymbol{\cdot} \hat{\boldsymbol{u}}_*)^2 \hat{\boldsymbol{u}}_*}{\dot{M}_p} \approx \frac{2Hu\rho_*u_*^2 \hat{\boldsymbol{u}}_*}{\dot{M}_p}
\end{align}
where $\dot{M}_p$ is the planetary mass-loss rate, and the velocity of the planetary gas is small compared to the stellar wind velocity. At small deviations from the planetary semi-major axis $a_p$, the centrifugal force, stellar gravity, and radial coriolis force approximately balance; therefore, the expulsion of the gas is driven by the impact of the stellar wind. 

The change in velocity of the planetary gas due to the radial acceleration of the stellar wind in the time $t_s \approx \frac{2\pi a_p}{u_\phi} \approx \frac{2\pi a_p}{u}$ required for the gas to wrap fully around the star is:
\begin{align}
du \approx \frac{4\pi H\rho_*u_*^2a_p}{\dot{M}_p}
\end{align}
The formation of a torus requires $du/u \lesssim 1$.  At small deviations from the planetary semi-major axis, the height of the tube is approximately $\frac{u}{\Omega}$, with $\Omega$ the angular velocity of the planet \citep{Owen2023}. Therefore:
\begin{align}
\frac{du}{u} \approx \frac{\dot{M}_*u_*}{\dot{M}_pu_k}
\end{align}
where $u_k$ is the Keplerian velocity of the planet, and we have assumed the stellar wind is spherically symmetric: $\rho_*u_*^2 = \frac{\dot{M}_*u_*}{4\pi a_p^2}$.

At the short orbital periods of planets undergoing extreme mass loss, the stellar wind velocity (assuming solar wind-like properties) is expected to be on the order of a few hundred km s$^{-1}$ \citep[e.g.,][]{Johnstone2015}, comparable to the planet’s Keplerian velocity. Therefore, we expect that a circumstellar torus forms when the planetary mass-loss rate exceeds that of the stellar wind. Sun-like stars are thought to have mass-loss rates ranging from $\sim 10^{11}-10^{14}$ g s$^{-1}$ \citep{Cohen2011, Wood2021}. Therefore, a planetary mass-loss rate exceeding this is necessary for a torus to form. These mass-loss rates are only typically achieved for young planets, or planets that are close to overflowing their Roche lobe. We note that the condition we find for the formation of a circumstellar torus is in line with the results of 3D hydrodynamics simulations \citep[][]{McCann2019, Macleod2022}.  

\subsubsection{Radiation pressure}

Radiation pressure from stellar Lyman-$\alpha$ photons also imparts a force on the escaping planetary gas. For Sun-like stars, stellar wind ram pressure is expected to dominate over radiation pressure \citep[e.g.][]{Khodachenko2019, Debrecht2020, Carolan2021a, Owen2023}. However, it may be important for different stellar types. The force per unit length on the planetary gas due to radiation pressure is:
\begin{align}
f_{rad} = \frac{HL_{\text{Ly}\alpha}}{2\pi a_p^2c}
\end{align}
where $L_{\text{Ly}\alpha}$ is the Ly$\alpha$ luminosity of the star and $c$ is the speed of light. The change in velocity of the planetary gas in the time required for the gas to wrap fully around the star is:
\begin{align}
\frac{du}{u} \approx \frac{L_{\text{Ly}\alpha}}{\dot{M}_pu_kc}
\end{align}
For a planet orbiting Sun-like star with a $L_{\text{Ly}\alpha} \sim 10^{28} \text{ erg s}^{-1}$, a mass-loss rate of $\sim 10^{11} \text{g s}^{-1}$ is required to stop radiation pressure from dispersing the planetary gas.  

\subsubsection{Stellar magnetic field}

If the planet undergoing mass loss is located within the Alfvén radius of the star, then the stellar magnetic field energy density exceeds the kinetic energy density of the stellar wind. In this regime, the magnetic forces imparted on the (ionized) outflowing planetary gas by the stellar magnetic field dominates over the force imparted by stellar wind ram pressure. Since close-in planets typically reside within the co-rotation radius of their host stars, the stellar magnetic field will extract angular momentum from the planetary gas. If the magnetic field is sufficiently strong, this angular momentum removal will cause the planetary material to be quickly accreted onto the star, and prevent the formation of a stable circumstellar torus.

Understanding this interaction is a complex problem that ultimately requires three-dimensional magnetohydrodynamic (MHD) simulations. Whilst there have been previous 3D MHD atmospheric escape simulations that have included stellar and planetary magnetic fields \citep[e.g.,][]{Matsakos2015, Carolan2021b, Khodachenko2021, Cohen2022, Presa2024}, these have either not been global simulations or not simulated the regime we are considering. However, we can gain some insight into the fate of the planetary gas by comparing its kinetic energy density to the magnetic energy density of the star. This approach is reminiscent of how the magnetospheric boundary
($r_m$) of an accreting object is often estimated, which is typically formulated by equating the stellar magnetic field energy with the kinetic energy of the spherically accreting material \citep[e.g.,][]{Elsner1977}:
\begin{align}
&\frac{B(r_m)^2}{8\pi} \sim \frac{\dot{M}}{4\pi r_m^2}\sqrt{\frac{2\text{G}M_*}{r_m}} 
\end{align}
where $B$ is the stellar magnetic field strength, $\dot{M}$ is the accretion rate of the material and G is the gravitational constant. The square-rooted term corresponds to the free-fall velocity of the gas, which is $\sqrt{2}$ greater than Keplerian velocity. Assuming a dipolar stellar magnetic field, such that:
\begin{align}
B(r) = B_*\left(\frac{R_*}{r}\right)^3
\end{align}
where $B_*$ is the stellar surface magnetic field strength and $R_*$ is the stellar radius, the radius of the magnetospheric boundary is estimated to be:
\begin{align}
r_m \sim \left(\frac{B_*^4R_*^{12}}{2\text{G}M_*\dot{M}^2}\right)^{\frac{1}{7}}
\end{align}
Although this geometry is not directly applicable to the case we examine in this paper, we note that if the accretion rate required to push the magnetospheric boundary radius inside the planet's semi-major axis exceeds the planetary mass-loss rate, magnetic stresses dominate the flow. In this regime, the planetary material would be strongly disrupted by the stellar magnetic field, and would likely be rapidly funneled onto the star, preventing the formation of a stable circumstellar torus. Therefore, a mass-loss rate of at least 
\begin{align}
\dot{M}_p \gtrsim 10^{11}\left(\frac{B_*}{1\text{ G}}\right)^2\left(\frac{R_*}{\text{R}_{\odot}}\right)^6\left(\frac{M_*}{\text{M}_{\odot}}\right)^{-\frac{1}{2}}\left(\frac{a_p}{0.03\text{ AU}}\right)^{-\frac{7}{2}} \text{ g s}^{-1}
\end{align}
is required for a circumstellar torus to form. We caution that this is a lower limit for the required mass-loss rate to form a circumstellar torus, as the stellar magnetic field lines may thread and torque the planetary gas at orbital distances greater than the aforementioned magnetospheric boundary \citep[]{Bessolaz2008}. 

Another quantity of interest is the ratio of the thermal pressure of the planetary gas to the magnetic pressure, termed plasma beta. 
\begin{align}
\beta = \frac{\rho c_s^2}{B^2/8\pi}
\end{align}
where $c_s$ is the sound speed of the planetary gas. When $\beta \gg 1$, the dynamics of the planetary gas is dominated by thermal motions, and we do not expect the stellar magnetic field to disrupt the formation of a circumstellar torus. The sound speed of photoevaporative outflows is $\sim$ 10 km s$^{-1}$, which is about an order of magnitude less than the Keplerian velocity at these orbital radii. Therefore, $\beta \gtrsim 1$, when $\dot{M}_p \gtrsim 10^{13}$ g s$^{-1}$.

\subsection{Geometry of the circumstellar gas}

Because the long-term radial evolution of the circumstellar gas is unknown, the long-term radial extent of the gas is uncertain. We shall assume that the gas is confined to a narrow ring around the planet, dictated by the initial orbits on which the gas is launched \citep[]{McCann2019, Owen2023}. Therefore, we fix the inner and outer edges of the gas as follows:
\begin{align}
&R_{\text{in}} = a_p - L_{\text{cor}}\\
&R_{\text{out}} = a_p + L_{\text{cor}}\\
&L_{\text{cor}} = \frac{u}{2\Omega} \approx \frac{c_s}{2\Omega}
\end{align}
where $a_p$ is the semimajoraxis of the planet, $u$ is the velocity at which the gas is launched from the planet, which is approximately the escape velocity, $\Omega$ is the planetary angular velocity and $L_\text{cor}$ is an estimate of the maximum distance that the gas travels radially from the planet before being deflected onto an azimuthal orbit around the star \citep{Owen2023}. 

The initial vertical extent of the planetary gas is approximately given by the maximum height that a gas parcel fired from the planet reaches under the influence of stellar gravity $\sim \frac{u}{\Omega}$. Over time, we expect the gas to settle into vertical hydrostatic equilibrium with scale height: 
\begin{align}
h = \frac{c_s}{\Omega}
\end{align}
To see that this is the case, consider the vertical sound crossing time of the gas $t_{sc}$:
\begin{align}
t_{\text{sc}} = \frac{u}{\Omega}\frac{1}{c_s} \approx \frac{1}{\Omega}
\end{align}
{\rc noting that $u \approx  u_{\text{esc}} \approx c_s$}. This is much less than the lifetime of the circumstellar gas, which by definition, lasts for many planetary orbital periods. For simplicity, we take the height of the circumstellar gas to be equal to its scale height (i.e. $H = h = \frac{c_s}{\Omega}$). {\rc For wavelengths at which the gas is highly optically thick, there may be significant absorption from gas above this height, therefore; in this scenario, our model underpredicts the total absorption from the gas.}

\subsection{Composition and ionization structure of the circumstellar gas}\label{sec:composition}

The circumstellar gas largely inherits its composition from the upper atmosphere of the planet from which it originates, resulting in H/He dominated gas. The metallicity of the gas is influenced by both the planet’s atmospheric metallicity, which is generally enriched in metals relative to solar composition \citep[e.g.,][]{Thorngren2016}, and the degree of metal fractionation in the outflow \citep[e.g.,][]{Hunten1987, Zahnle1990}. Consequently, the metallicity of the circumstellar gas can span a broad range, from sub-solar to super-solar, with an upper limit set by the atmospheric metallicity of the planet.

The upper atmosphere of the planetary gas is highly irradiated due to its close proximity to its host star. Therefore, the gas that forms the torus is expected to be hot ($\sim 10^4$ K) and composed of atoms, ions, and free electrons \citep[e.g.,][]{Murray-Clay2009}. As the torus is optically thin to XUV radiation (since a H/He dominated gas will be optically thin to X-rays if it is optically thin to EUV photons), we expect the circumstellar gas to become progressively more ionized until it reaches photoionization-recombination equilibrium. The timescale for a species to reach photoionization-recombination equilibrium is approximately the recombination timescale:
\begin{align}
t_r = \frac{1}{\alpha_\text{A}n_e} 
\end{align}
where $\alpha_\text{A}$ is the case $\text{A}$ recombination coefficient of the species of interest and $n_e$ is the electron number density. At an electron density of $n_{e} = 10^{7} \text{ cm}^{-3}$ (somewhat less than the expected density of the circumstellar gas; see Section~\ref{sec:density}), the recombination time for atomic hydrogen, using $\alpha_\text{A} = 4.18 \times 10^{-13} \text{ cm}^3\text{s}^{-1}$ at $10^4$ K, is $\sim 3 \text{ days}$, which is much shorter than the expected lifetime of the circumstellar gas; therefore, the hydrogen population in the circumstellar gas will be in photoionization-recombination equilibrium. Typical recombination rates for other species range from $10^{-13}-10^{-11} \text{ cm}^3\text{s}^{-1}$ \citep{Bryans2006} therefore we expect all species to be in photoionization-recombination equilibrium. In photoionization-recombination equilibrium, the circumstellar gas maintains thermal equilibrium at approximately $10^4$ K, where heating from photoionization is balanced by radiative cooling, dominated primarily by Ly$\alpha$ and metal line cooling \citep[e.g.,][]{Murray-Clay2009, Owen2016, Huang2023}.  

Since the circumstellar gas has a relatively low density (n $\lesssim 10^9$ cm$^{-3}$), we expect that detectable absorption will primarily come from strong absorption lines originating from transitions out of highly populated electronic states. At a temperature of $10^4$ K, the gas is too cold for excited states to be significantly populated. Therefore, our analysis focuses chiefly on observational signatures arising from ground-state absorption lines. Accordingly, we compute the state of hydrogen and metals in photoionization-recombination equilibrium, assuming that all atoms and ions are in the ground state. An exception to this is helium, where we are interested in the He\,\textsc{i} 10830 \AA~absorption line, which arises from the 2$^3\text{s} \rightarrow  2^3\text{p}$ electronic transition. This absorption line has been frequently observed in exoplanet outflows \citep[e.g.,][]{Spake2018, Allart2018}, due to the abundance of He\,\textsc{i} in these flows and the fact that the metastable 2$^3\text{s}$ state is readily populated through the recombination cascade \citep[e.g.][]{Osterbrock2006, Oklopvcic2018}.
    
As the circumstellar gas is optically thin to XUV radiation, the photoionization rate of a given species remains approximately constant throughout the torus. The optically thin ionization rate for a species of atomic number $Z$ and ionization state $i$ is given by:
\begin{align}
    \Gamma(Z_i) = \int_{\nu_0(Z_i)}^{\infty}\frac{L_{\nu}}{4\pi r^2h\nu}\sigma_{\nu}(Z_i)d\nu
\label{eq:gamma}
\end{align}
where $L_{\nu}$ is the stellar luminosity, $\nu_0(Z_i)$ is the photoionization threshold frequency of the species and $\sigma_\nu(Z_i)$ is frequency dependent photoionization cross section. Given a stellar spectrum, the photoionization rate for a species is calculated by numerically integrating Equation~\ref{eq:gamma}. For all atoms and ions in the ground state, the frequency-dependent photoionization cross sections are taken from the analytic fits provided by \citet{Verner1996}. For the He\,\textsc{i} $2^3\text{s}$ state, the photoionization cross section is taken from tabulated data provided by \citet{Norcross1971}.

\subsubsection{Hydrogen}

In photoionization-recombination equilibrium, the equations governing the populations of atomic and ionized hydrogen are:
\begin{align}
    &X_{\text{H\,I}}\Gamma(\text{H\,\textsc{i}}) - n_{e}X_{\text{H\,II}}\alpha_{\text{A}}(\text{H\,}\textsc{i}) = 0\\
    &X_{\text{H\,I}} + X_{\text{H\,II}} = 1
    \label{ionisation_eq}
\end{align}
where $X_{\text{H\,I}}$ and $X_{\text{H\,II}}$ are the fraction of hydrogen that is neutral and ionized respectively, $\Gamma(\text{H\,}\textsc{i})$ is the ionization rate of neutral hydrogen, and $\alpha_{\text{A}}(\text{H }\textsc{i})$ is the Case A recombination coefficient for hydrogen. Since the composition of the circumstellar gas is hydrogen dominated, we approximate the number density of free electrons to be equivalent to the number density of ionized hydrogen: $n_e \approx X_{\text{H\,II}}n_{\text{H}}$, where $n_{\text{H}}$ is the total number density of hydrogen (netural and ionized). Therefore, the fraction of hydrogen that is ionized is:
\begin{align}
    X_{\text{HII}} = \frac{\sqrt{\Gamma(\text{H}\,\textsc{i})^2 + 4n_{\text{H}}\alpha_{\text{A}}(\text{H}\,\textsc{i})  \Gamma(\text{H}\,\textsc{i})} - \Gamma(\text{H}\,\textsc{i})}{2n_{\text{H}}\alpha_{\text{A}}(\text{H}\,\textsc{i}) }
\end{align}

\subsubsection{Helium}

{\rc Neutral helium atoms have two possible total spin eigenstates (S): a singlet state (S=0) and a triplet state (S=1). Radiative transitions from the lowest energy triplet state ($2^3$s) to the ground state ($1^1$s) are spin-forbidden and are therefore strongly suppressed. Because of this slow radiative transition, a significant population of helium in the circumstellar gas may be in the $2^3$s state. To determine the population of helium atoms in the $2^3$s state, we employ a simplified version of the network presented in \citet{Schulik2025}, based on earlier models developed by \citet{Oklopvcic2018, Allan2024}. In \citet{Schulik2025}, the populations of helium in the $1^1$s, $2^3$s, $2^1$s, $2^1$p, He \textsc{ii}, He \textsc{iii} states are computed. However, they showed that the $2^1$s and $2^1$p states are highly depleted and therefore that collisional excitation from the $2^3$s state to one of these states can effectively be treated as a transition to the ground state ($1^1$s). Therefore, in this work, we only calculate the helium populations of the $1^1$s, $2^3$s, $2^1$s, $2^1$p. 

In kinetic equilibrium, the helium state populations are found by balancing the processes (photoionization, recombination, collisional excitation/de-excitation and radiative decay) that populate and de-populate each state each state. The fraction of helium in each state is found by solving the resulting set of coupled equations, each representing the kinetic balance of a given state.}:

\begin{align}
\label{eq:singlet_eq}
\begin{split}
&X_{\text{He\,II}}n_e\alpha_\text{A}(\text{He\,\textsc{i} } n^1\ell) + X_{\text{He\,I }2^3\text{s}}A_{31} + X_{\text{He\,\textsc{I} }2^3\text{s}}q_{31a}n_e\\ 
&+ X_{\text{He\,\textsc{I} }2^3\text{s}}q_{31b}n_e + X_{\text{He\,\textsc{I} }2^3\text{s}}q_{31c}n_e + X_{\text{He\,\textsc{I} }2^3\text{s}}Q_{31}n_{\text{H\,\textsc{i}}}\\
&- X_{\text{He\,\textsc{I} }1^1s}\Gamma({\text{He\,\textsc{i} }1^1\text{s}})  - X_{\text{He\,\textsc{I} }1^1\text{s}}q_{13}n_e = 0\\ 
\end{split}
\end{align}
\begin{align}
\label{eq:triplet_eq}
\begin{split}
&X_{\text{He\,II}}n_e\alpha_\text{A}(\text{He\,\textsc{i} } n^3\ell) - X_{\text{He\,\textsc{I} }2^3\text{s}}A_{31} - X_{\text{He\,\textsc{I} }2^3\text{s}}q_{31a}n_e\\ 
&- X_{\text{He\,\textsc{I} }2^3\text{s}}q_{31b}n_e - X_{\text{He\,\textsc{I} }2^3\text{s}}q_{31c}n_e - X_{\text{He\,\textsc{I} }2^3\text{s}}Q_{31}n_{\text{H\,\textsc{i}}}\\ 
&- X_{\text{He\,\textsc{i} }2^3\text{s}}\Gamma(\text{He\,\textsc{i} }2^3\text{s})  + X_{\text{He\,\textsc{i} }1^1\text{s}}q_{13}n_e = 0 
\end{split}
\end{align}
\begin{align}
X_{\text{He\,II}}\Gamma(\text{He\,\textsc{ii}}) - X_{\text{He\,III}}\alpha_\text{A}(\text{He\,\textsc{ii}}) = 0
\end{align}
\begin{align}
X_{\text{He\,\textsc{I} } 1^1\text{s}} + X_{\text{He\,\textsc{I} } 2^3\text{s}} + X_{\text{He\,\textsc{II}}} + X_{\text{He\,\textsc{III}}} = 1 
\end{align}
where $X_{\text{He\,\textsc{I} } 1^1\text{s}}$, $X_{\text{He\,\textsc{I} }2^3\text{s}}$, $X_{\text{He\,II}}$, $X_{\text{He\,III}}$ denote the fraction of helium in each state, $q_{ij}$ denote electronic transitions due to collisions with electrons, $Q_{ij}$ denote electronic transitions due to collisions with hydrogen atoms, {\rc $\Gamma(Z_i)$ denote photoionizations, and $\alpha_{\text{A}}(Z_i)$ denote case A recombinations. The values of these coefficients are given in Table~\ref{tab:Helium State Processes}. The dominant pathway for populating the metastable state is recombination from He\,\textsc{ii} \citep[e.g.,][]{Osterbrock2006, Oklopvcic2018}. The dominant depopulation pathways are collisional excitation to the $2^1\text{s}$ state and ionization into the He\,\textsc{ii} state \citep[e.g.,][]{Oklopvcic2018, Schulik2025}.}

\begin{table*}
\centering
 \begin{threeparttable}
 \caption{The collisional and radiative processes considered in our modelling of the helium state populations.}
 \label{tab:Helium State Processes}
  \begin{tabular}{l l l l}
  Populates & Depopulates & Rates & References\\
  & & \textit{Recombination} &\\
  He\,\textsc{i} $1^1$s & He\,\textsc{ii} & $\alpha(\text{He\,}\textsc{i } n^1\text{l}) = 2.20 \times 10^{-10}T^{-0.678} - 2.72 \times 10^{-10}T^{-0.778}$ & a\\
  He\,\textsc{i} $2^3$s & He\,\textsc{ii} & $\alpha(\text{He\,}\textsc{i } n^3\text{l}) = 2.72 \times 10^{-10}T^{-0.778} $& a\\
  He\,\textsc{ii} & He\,\textsc{iii} & $\alpha_\text{A}(\text{He II}) = 1.82 \times 10^{-10}\sqrt{\frac{T}{10.17}}\left(1+\sqrt{\frac{T}{10.17}}\right)^{0.251}\left(1+\sqrt{\frac{T}{2.79\times10^6}}\right)^{1.749}$ & b,c\\
  & & \textit{Collisional (de-)excitation} &\\
  He\,\textsc{i} $1^1$s & He\,\textsc{i} $2^3$s & $q_{31\text{a}} = 3.78\times10^{-5}T^{-0.690}\text{exp}(\frac{-1.02 \times 10^4}{T})$ & b,d\\
  He\,\textsc{i} $1^1$s & He\,\textsc{i} $2^3$s & $q_{31\text{b}} = 1.52\times10^{-6}T^{-0.438}\text{exp}(\frac{-1.69 \times 10^4}{T})$ & b,d\\
  He\,\textsc{i} $1^1$s & He\,\textsc{i} $2^3$s & $q_{31\text{c}} = 2.60\times10^{-7}T^{-0.531}\text{exp}(\frac{-5.43 \times 10^2}{T})$ & b,d\\
  He\,\textsc{i} $2^3$s & He\,\textsc{i} $1^1$s & $q_{13} = 7.79\times10^{-7}T^{-0.531}\text{exp}(\frac{-2.31 \times 10^5}{T})$ & b,d\\
  He\,\textsc{i} $1^1$s & He\,\textsc{i} $2^3$s & $Q_{31}= 5.0 \times 10^{-10}$ & e\\
  & & \textit{Radiative Decay} & \\
  He\,\textsc{i} $1^1$s & He\,\textsc{i} $2^3$s & $A_{31} = 1.272 \times 10^{-4}$ & f\\
  \end{tabular}
  \begin{tablenotes}
    \item The letters correspond to the following references: a-\citet{Benjamin1999}, b-\citet{CHIANTI1, CHIANTI10}, c-\citet{Badnell2006},d-\citet{Schulik2025}, e-\citet{Roberge1982}, f-\citet{Drake1971}. The collisional transition rates are fitting functions from \citet{Schulik2025} using data from \citet{CHIANTI1, CHIANTI10}. 
  \end{tablenotes}
  \end{threeparttable}

\end{table*}

\subsubsection{Metals}

In photoionization-recombination equilibrium, the fraction of a species (of atomic number Z) in each ionization state is described by a system of equations:
\begin{align}
    &X_{i}\Gamma(Z_i) - n_eX_{i+1}\alpha_{\text{A}}(Z_{i}) = 0 \quad \quad i = \{0,1, ... ,Z-1\}\\
    &\sum_{i=0}^{Z} X_{i} = 1
    \label{eq:ionisation_Zeq}
\end{align}
where $X_{i}$ denotes the ionization fraction of the species in ionization state $i$, where $i$ ranges from $0$ (neutral) to $Z$ (fully ionized). For all metals, we use the case A recombination coefficients provided in the \textsc{CHIANTI} database \citep{CHIANTI1, CHIANTI10}. Approximating the electron number density as equal to the number density of ionized hydrogen decouples the ionization equilibrium of different species. This allows each species' ionization equilibrium to be solved independently through matrix inversion, using an LU decomposition routine from \textsc{LAPACK} \citep{LAPACK}.

\subsection{Density of the circumstellar gas}
\label{sec:density}

As discussed previously, the density structure of the circumstellar gas is uncertain due to the incomplete understanding of the physical mechanisms that shape the circumstellar torus. Therefore, we assume the density is constant, and is limited by the condition that the gas between the star and the planet remains optically thin to EUV radiation:
\begin{align}
\tau_{\text{EUV}} = N_{\text{H\,I}}\sigma_{\text{H\,I}} \leq 1
\end{align}
where $N_{\text{H\,I}} = 2n_{\text{H\,I}}L_{\text{cor}}$ is the column density of neutral hydrogen and $\sigma_{\text{H\,I}} = 6.3 \times 10^{-18}$ cm$^2$ is the threshold photoionization cross section of neutral hydrogen. To satisfy this condition, the total hydrogen density is:
\begin{align}
n_{\text{H}} \leq \frac{1}{L_{\text{cor}}\sigma_{\text{H\,I}}}\left(1 + \sqrt{\frac{L_{\text{cor}}\sigma_{\text{H\,I}}\Gamma(\text{H}\,\textsc{i})}{\alpha_\text{A}(\text{H}\,\textsc{i})}}\right) 
\end{align}
For a typical $L_{\text{cor}} \sim 10^{10}$ cm and $\Gamma(\text{H}\,\textsc{i}) \sim 10^{-4} \text{ s}^{-1}$, this gives a maximum hydrogen density of $\sim 5\times 10^8 \text{ cm}^{-3}$, and a maximum mass of the circumstellar torus of $M_r \sim 5 \times 10^{18}$ g. The timescale to build a torus of this mass is:
\begin{align}
t_m \sim 60 \left(\frac{M_r}{5 \times 10^{18}\text{ g}}\right)\left(\frac{\dot{M}_p}{10^{12} \text{ g}}\right)^{-1} \text{ days}
\end{align}
This timescale is much shorter than the age, mass loss timescale and orbital decay timescale of short period planets. Therefore, the assumption that the circumstellar torus will have reached this EUV-limited maximum density is reasonable. 

\section{Synthetic Observations}\label{sec:synthetic_transit}

We employ a ray-tracing scheme to compute the attenuation of stellar light by the circumstellar gas. We construct a Cartesian coordinate system $(x,y,z)$, with the origin at the center of the stellar disk, $\mathcal{D}$. The $x$ and $y$ axes lie in the plane of the stellar disk and the $z$-axis points towards the observer. The occulted stellar intensity is:
\begin{align}
I_\nu = \iint_\mathcal{D} I_{*,\nu}(x,y)e^{-\tau_{\nu}(x, y)} dxdy 
\end{align}
where $I_{*,\nu}$ is the unocculted intensity and $\tau_{\nu}$ is the optical depth of the circumstellar gas at frequency $\nu$. For simplicity we treat the stellar disk as spatially homogeneous such that:
\begin{align}
I_{*,\nu}(x,y) = I_{*,\nu}
\end{align}
In reality, the surface brightness of a star is inhomogenous, especially in chromospheric emission lines, which are dominated by emission from active regions. However, the large occulting area of the circumstellar gas means that it will likely cover both a large amount of active regions and non-active regions so that it's total  effect is less affected by surface inhomogenity.

The optical depth of the circumstellar torus is: 
\begin{align}
\tau_{\nu}(x,y) = \int_{0}^{\infty}\sum_{s} n_{s}\sigma_{s,\nu}(u_{z}, u_{\text{turb}}, T) \ dz
\label{eq:optical_depth}
\end{align}
where $n_s$ is the number density of species $s$ and $\sigma_{s,\nu}$ is the species absorption cross-section at frequency $\nu$, which depends on {\rc the temperature of species, the bulk (line of sight) velocity, $u_z$ and turbulent velocity, $u_{\text{turb}}$ of the gas. We assume that line broadening due to the distribution of turbulent velocities along the line of sight is Gaussian with standard deviation of  $\frac{u_{\text{turb}}}{\sqrt{2}c}$ \citep[e.g.,][]{Struve1934}.} The total absorption cross-section of a species is sum of the individual electron transition cross-sections. The absorption cross-section corresponding to a transition is given by:   
\begin{align}
    \sigma_{\nu} = \frac{\pi e^2}{m_{e}c}f\Phi(u_{z}, u_{\text{turb}}, T)
\end{align}
where $e$ and $m_e$ is the charge and mass of the electron, $c$ is the speed of light, $f$ is the oscillator strength of the transition and $\Phi$ is the Voigt line profile. The Gaussian part of the Voigt profile has a standard deviation of $\sigma_N$, the Lorentzian part has a half-width half-maximum of $\gamma$, and the line centre has been Doppler shifted to $\nu$. These are given below:
\begin{align}
    \sigma_N = \nu_0\sqrt{\frac{k_bT}{m_{s} c^2} + \frac{u_{\text{turb}}^2}{2c^2}} ,\quad \gamma = \frac{A}{4\pi}, \quad \nu = \nu_0\left(1 - \frac{u_{z}}{c}\right)
\end{align}
where $m_s$ is the mass of species, $k_b$ is the Boltzmann constant and $A$ is the Einstein A coefficient of the transition. For each species and line, the oscillator strengths and Einstein A coefficients are taken from the CHIANTI atomic database \citep{CHIANTI1, CHIANTI10}. 

The circumstellar torus model requires eight input parameters listed in Table~\ref{tab:Model_Parameters}. 

\begin{table}
  \caption{The free parameters of the circumstellar torus model.}
  \begin{tabular}{p{2.25cm}p{5.25cm}}
   Stellar Parameters & Planet/Torus Parameters\\
   \hline
   Stellar Mass ($M_*$) & Planet semi-major axis ($a_p$)\\
   Stellar Radius ($R_*$) & Gas Temperature ($T_r$)\\
   Stellar SED ($L_{*,\nu}$) & Gas Turbulent Velocity ($u_\text{turb}$)\\
   & Torus Inclination ($i_r)$\\
   & Torus Optical Depth to EUV radiation ($\tau_{\text{EUV}})$\\
   & Torus Composition\\
   \end{tabular}
   \label{tab:Model_Parameters}
\end{table}

\section{Methods}\label{sec:methods}

In this section, we model the effect that circumstellar tori have on the observed properties of the stellar He\,\textsc{i} 10830 \AA~and Ca\,\textsc{ii} H \& K lines. To characterize the impact of these tori over a range of system parameters, we generate an ensemble of stellar systems and apply absorption from modeled circumstellar tori. The stellar properties in our sample are internally consistent, ensuring that individual stars are realistic. However, the sample is not intended to match any specific observed stellar population; instead, the sample is designed to broadly span the parameter space where circumstellar tori may form. For example, our sample contains an equal number of F, G, and K stars. 

\subsection{Constructing the stellar sample}

We construct a stellar sample consisting of FGK-type stars with masses ranging from 0.8 to 1.2 M$_\odot$. This sample is divided into three equal-sized sub-samples based on stellar type: K-type stars with masses between 0.85–0.9 M$_\odot$, G-type stars with masses between 0.9–1.1 M$_\odot$ and F-type stars with masses between 1.1–1.25 M$_\odot$. The masses of the stars in each sample is drawn from a uniform distribution. From the stellar mass, the corresponding radius and bolometric luminosity of the star is calculated using the equations (from \citealt{Demircan1991}):
\begin{align}
&\frac{R_*}{R_\odot} = 1.06 \left(\frac{M_*}{M_\odot}\right)^{0.945}\\
&\frac{L_{*,\text{bol}}}{L_{\odot,\text{bol}}} = 1.02 \left(\frac{M_*}{M_\odot}\right)^{3.92}
\end{align} 

To generate a realistic high-energy spectral energy distribution (SED) for each star, we first take a representative template spectrum for the stellar type. The template stars used are as follows: late-F: HD 108147, early-G: HD 149026, solar, late-G: TOI-193, early-K: HD 97658. The stellar data was downloaded from \textsc{p-winds}\footnote{https://github.com/ladsantos/p-winds}, which obtained the majority of the stellar data from the MUSCLES library \citep[e.g.,][]{France2016, Youngblood2016, Loyd2016}, which combines observed and theoretical data to produce high-energy spectral templates (from NUV to X-ray) for various stellar types. The solar spectrum used is the 2008 Whole Heliosphere Interval (WHI) Solar Irradiance Reference Spectra (SIRS) \citep{Woods2009} \footnote{Data accessed via the LASP Interactive Solar Irradiance Datacenter (LISIRD) (https://lasp.colorado.edu/lisird/)}. The HR 108147 spectrum was obtained from combining observations from the X-exoplanets database \citep[]{Sanz-Forcada2010} with PHOENIX atmosphere models from the NUV \citep{Allard1995, Husser2013}. 

To make a high-energy spectrum for the sample star, we draw the level of activity of the sample star, and then modify the high-energy template accordingly. A star's activity is not uniquely defined, as it can be quantified through a variety of indicators such as chromospheric and coronal emission or photometric variability. In this work, we characterize stellar activity using three complementary metrics: (1) the X-ray luminosity in the 5.17–124 \AA~band, corresponding to the range observed by the \textit{ROSAT} satellite, divided by the bolometric luminosity, $\left(\frac{L_\text{X}}{L_{\text{bol}}}\right)$; (2) the log\,$R^{'}_{\text{HK}}$ index, which traces chromospheric Ca\,\textsc{ii} emission; and (3) the equivalent width of the He\,\textsc{i} 10830 \AA~line. As these quantities are not independent, they must be sampled jointly to maintain physically realistic correlations.

To jointly sample these quantities, we proceed as follows. We take the subset of stars from the \citet{Smith2016} catalog that have measured values for X-ray luminosity, the equivalent width of the He\,\textsc{i} 10830 \AA~line, and the log\,$R^{'}_{\text{HK}}$. We further restrict this sample to stars with a B–V colour greater than 0.47 and a $\frac{L_\text{X}}{L_{\text{bol}}} \leq -4 $, consistent with the regime where the correlation described in Section 2 holds. This yields a final sample of 35 stars. From this subset, we construct a Kernel Density Estimator (KDE) to randomly draw the joint $\frac{L_\text{X}}{L_{\text{bol}}}$,  He\,\textsc{i} 10830 \AA~equivalent width and log\,$R^{'}_{\text{HK}}$ for the synthetic stellar sample.  

To modify the template star's high-energy SED based on these drawn quantities, we divide the spectrum into four energy bands: X-ray, EUV, FUV, and NUV, each of which scales differently with stellar activity. We estimate the flux of the sample star in each of these bands from the drawn X-ray flux and log\,$R^{'}_{\text{HK}}$, using the empirically derived scaling relations below:
\begin{itemize} 
\item EUV: Estimated from the X-ray flux using the power-law relation from \citet{King2018}, specifically using the parameters from \#1 in their Table 1. 
\item FUV and NUV: Estimated from log\,$R^{'}_{\text{HK}}$ using the relations provided in \citet{Findeisen2011}. These equations require B and V stellar magnitudes as a function of stellar mass, which we also adopt from \citet{Findeisen2011}. 
\end{itemize}
Because the energy bands lack universally defined wavelength ranges, we adopt two distinct sets of definitions, as detailed in Table~\ref{tab:Energy_Bands}. The comparison range refers to the wavelength interval over which the scaling relations we use to estimate the sample stars EUV, FUV, and NUV from its drawn X-ray flux and log\,$R^{'}_{\text{HK}}$ are valid. We use this range to calculate how much the template spectrum needs to be scaled in an energy band to match these  estimated fluxes. The scaling range defines the wavelengths that get scaled by the factor. 

\begin{table*}
  \caption{The definitions of the energy ranges of different XUV bands used in this work.}
  \begin{tabular}{p{3cm}p{3cm}p{3cm}p{3cm}p{3cm}}
   Band & Scaling Range (\AA) & Comparison Range (\AA) & Estimation Method Ref. & Relevant Satellite\\
   \hline
   X-ray & <100 & 5.17-124 & N/A & \textit{ROSAT}\\
   EUV & 100-1200 & 124-912 & \citet{King2018} & N/A\\
   FUV & 1200-1780 & 1350-1780 & \citet{Findeisen2011} & \textit{GALEX} FUV\\
   NUV & 1780-2830 & 1780-2830 & \citet{Findeisen2011} &  \textit{GALEX} NUV\\
   \end{tabular}
   \label{tab:Energy_Bands}
\end{table*}

\subsubsection{Intrinsic stellar lines}

To compute the modified He\,\textsc{i} 10830 \AA~equivalent width and the log\,$R_{\text{HK}}'$ due to an absorbing circumstellar torus, we need to consider the shape of the stellar He\,\textsc{i} 10830 \AA~and Ca\,\textsc{ii} H \& K lines. Detailed modelling of these stellar lines is beyond the scope of this work; instead, we adopt simplified, analytical representations that capture their essential features. 

\begin{figure}
\centering
\includegraphics[width=\columnwidth]{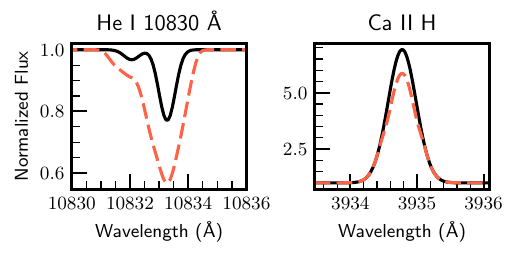}
\caption[]{{\rc Model He 10830 \AA~and Ca\,\textsc{ii} H stellar profiles for a star of $M_* = 0.92\text{M}_{\odot}$ with log\,$R_{\text{HK}}'$ = -4.89 and He\,\textsc{i} 10830 \AA~EW = 170 m\AA. The solid black lines shows the intrinsic stellar line profiles, and the dashed red line shows the lines modified by an obscuring circumstellar torus with a radius of 0.03 AU.}}    
\label{fig:template_lines}
\end{figure}

We model the shape of stellar He\,\textsc{i} 10830 \AA~absorption line to be a sum of three Voigt profiles, each centered on one of the triplet transitions. The relative strengths of the triplet components are set according to their oscillator strengths, so that the equivalent width of the line is determined by the temperature and a scaling factor. This approach assumes that the chromospheric gas forming the stellar line is optically thin, which is justified given the generally weak absorption observed. Thermal broadening is set by a characteristic chromospheric/transition region temperature of $2.5 \times 10^4$ K, and we neglect any contribution from rotational broadening. The amplitude of the scaling factor is calculated such that the total equivalent width of the He\,\textsc{i} 10830 \AA~line matches the value drawn for the considered star. An example stellar line is shown in Figure~\ref{fig:template_lines}. 

The Ca\,\textsc{ii} H \& K lines of FGK stars typically show emission cores from chromospheric emission inside of broad photospheric absorption wings. {\rc The circumstellar gas predominantly absorbs in the core of these Ca\,\textsc{ii} H \& K lines, therefore it is only necessary to model the core of these lines. We model the core of the Ca\,\textsc{ii} H \& K lines to comprise of two components: a constant photospheric background and a chromospheric emission component that has the shape of a Voigt profile, centred on the rest wavelength of the H or K transition. Our simple model does not include the central reversal in the emission core, as seen in the Sun. The central reversal is narrow (see Figure~\ref{fig:he_ca_correlation}) and usually contributes a small fraction of the total chromospheric emission, and therefore this omission has a minor effect.} Thermal broadening at a temperature of $2.5 \times 10^4$ K is not sufficient to broaden the lines to those observed \citep[e.g.,][]{Wilson1957}. Therefore, we artificially broaden the lines to have a full width half maximum (FWHM) of 0.45 \AA, equivalent to that of the Sun's Ca\,\textsc{ii} K line \citep[][]{Wilson1957, Stencel1977}. An example line is shown in Figure~\ref{fig:template_lines}.

Our aim is to assess how absorption from circumstellar gas affects the measured log\,$R_{\text{HK}}'$. To do this, we first determine how changes in the Ca\,\textsc{ii} H \& K lines modify the S-index, which quantifies the ratio of chromospheric emission in these lines relative to the nearby continuum. The S-index is defined as \citep{Vaughan1978, Middelkoop1982, Noyes1984}:
\begin{align}
S = \alpha\frac{N_\text{H} + N_\text{K}}{N_\text{V} + N_\text{R}} ,
\end{align}
where $N_\text{H}$ and $N_\text{K}$ are the fluxes integrated over $\sim1$ \AA~triangular bandpasses centered on the H and K lines, respectively, and $N_\text{V}$ and $N_\text{R}$ are the fluxes measured in two nearby continuum reference bands. $\alpha$ is scaling constant. In our simple model, the observed flux in the Ca\,\textsc{ii} H \& K lines is:
\begin{align}
N_{\text{H}} + N_{\text{K}} = \int_{\mathcal{H}} (f_{\text{H}}\Phi + g_\text{H})\,d\lambda + \int_{\mathcal{K}} (f_\text{K} \Phi + g_\text{K})\,d\lambda 
\end{align}
where $f_\text{H}$, $f_\text{K}$ and $g_\text{H}$ and $g_\text{K}$ are the amplitudes of the chromospheric component and photospheric component of the H and K lines respectively, and $\Phi$ is the Voigt profile.  The integrals are evaluated over the respective $\sim1$ \AA~bandpasses, labeled $\mathcal{H}$ and $\mathcal{K}$, centered on the H and K lines.

The S-index can be split into a photospheric and chromospheric part: $S = S_{\text{phot}} + S_{\text{chr}}$, where:
\begin{align}
&S_{\text{phot}} = \frac{\alpha}{N_\text{V}+N_\text{R}}\left[\int_{\mathcal{H}}  g_\text{H}\,d\lambda + \int_{\mathcal{K}} g_\text{K}\, d\lambda\right]\\
&S_{\text{chr}} = \frac{\alpha}{N_\text{V} + N_\text{R}} \left[ \int_{\mathcal{H}} f_{\text{H}}\phi\,d\lambda + \int_{\mathcal{K}} f_\text{K}\phi \,d\lambda \right] 
\end{align}
To estimate the absorption, we assume that the stellar profiles of the Ca\,\textsc{ii} H and K lines have equal amplitudes, which implies that the flux ratio $\frac{N_\text{H}}{N_\text{K}} \approx 1$. \citet{Noyes1984} demonstrated that this approximation holds reasonably well, with their stellar sample exhibiting minimal differences between the two lines. Based on this, we adopt the simplification $f_{\text{H}}=f_\text{K}=f$ and $g_\text{H}=g_\text{K}=g$. 

The modification of the observed S-index due to absorption by a circumstellar torus is then given by: 
\begin{align*}
S_{\text{torus}} &= \frac{\alpha}{N_V+N_R}\left[\int_{\mathcal{H}} (f\phi(\lambda) + g)I_\text{H}(\lambda)\,d\lambda \right.\\ 
&+ \left. \int_{\mathcal{K}} (f\phi(\lambda) + g)I_\text{K}(\lambda)\,d\lambda\right]\\
&= \frac{S_\text{phot}}{\int_{\mathcal{H}} d\lambda + \int_{\mathcal{K}} d\lambda}\left[\int_{\mathcal{H}} I_\text{H}(\lambda)\,d\lambda + \int_{\mathcal{K}} I_\text{K}(\lambda)\,d\lambda\right]\\
& + \frac{S_{\text{chr}}}{\int_{\mathcal{H}} \phi d\lambda + \int_{\mathcal{K}} \phi(\lambda) d\lambda}\left[\int_{\mathcal{H}} \phi(\lambda) I_\text{H}(\lambda)\,d\lambda + \int_{\mathcal{K}} \phi I_\text{K}(\lambda)\,d\lambda\right]
\end{align*}
where $I_{\text{H}}$ and $I_{\text{K}}$ is the absorption in the H and K lines from the circumstellar gas. To calculate this, we need to estimate the separate photospheric and chromospheric contribution to the S-index. The photospheric contribution $S_\text{phot}$ can be estimated as a function of B–V color using the empirical relation provided by \citet{Hartmann1984}. Finally, the modified S-index, $S_{\text{torus}}$, is converted to a modified log\,$R_{\text{HK}}'$ using the method of \citet{Noyes1984}. 

\subsection{Circumstellar torus properties}\label{sec:rings}

For each system, we generate a circumstellar torus using the model described in Section~\ref{sec:model}. This model requires a set of input parameters, which are listed in Table~\ref{tab:Model_Parameters}. Here, we describe our choices for these parameters in our sample. The orbital period of the planet, whose mass loss gives rise to the circumstellar torus, is drawn from a uniform distribution between 1 and 3 days, consistent with the typical periods of hot Jupiters \citep{Dawson2018}. We assume an impact parameter of zero for the circumstellar torus. Since the scale height of the circumstellar gas is much larger than the planetary radius and we focus on transiting or near-transiting systems, this is a reasonable approximation. 

As discussed in Section~\ref{sec:model}, the timescale for the gas density in the torus to become marginally optically thick to EUV radiation is much shorter than the timescale over which the planet’s properties evolve. Therefore, we assume that the majority of observed tori will have reached this density; therefore, for the modelled circumstellar tori, we fix the optical depth to EUV at a value of 2 (one between the planet and the star and one on the far side of the planet). Finally, we also set the gas to have a metallicity of five-times solar composition, consistent with observations of the atmospheres of giant planets \citep{Thorngren2016}. The temperature of the gas in the torus is fixed at $10^4$ K, expected for a gas in photoionization-recombination equilibrium (see Section~\ref{sec:composition}) 

\section{Results}\label{sec:he_vs_ca}

\begin{figure*}
\centering
\includegraphics[width=\textwidth]{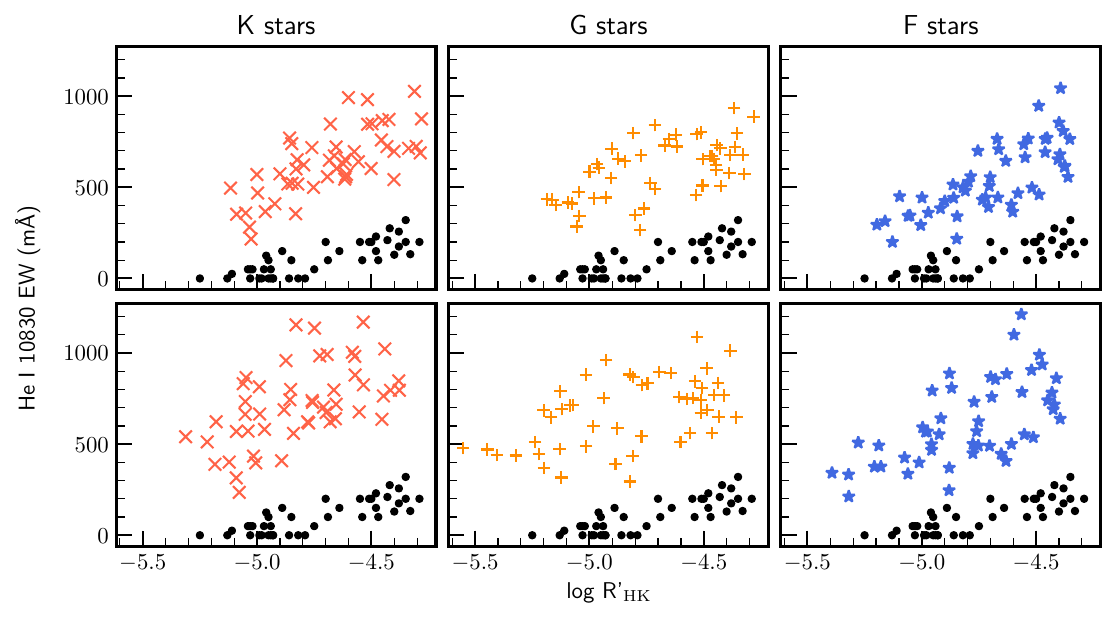}
\caption[Helium 10830 \AA~EW vs log $R^{'}_{HK}$ for stars hosting circumstellar disks]{{\rc The predicted He\, 10830 \AA~EW and log\,$R_{\text{HK}}'$ for the synthetic sample of FGK stars hosting circumstellar tori (blue stars, yellow pluses, red crosses respectively). For comparison, observed stars (without circumstellar tori) from the \citet{Smith2016} sample are plotted as black dots. The top panel shows models in which the gas is assumed to have no turbulent velocity ($u_{\text{turb}} = 0$), while the bottom panel shows models with $u_{\text{turb}} = 0.5c_s$.}}
\label{fig:circumstellar_comparison}
\end{figure*}

{\rc In Figure~\ref{fig:circumstellar_comparison} we show the calculated He\, 10830 \AA~EW and log\,$R_{\text{HK}}'$ for the sample of stars hosting circumstellar tori generated in Section~\ref{sec:methods}. The top panel shows the case where the circumstellar tori have $u_{\text{turb}} = 0$. The bottom panel shows the case where the circumstellar tori have $u_{\text{turb}} = 0.5c_s$. In both cases, there is a significant enhancement in the strength of the He\,\textsc{i} 10830 \AA~absorption line, which leads to a clear separation between the two populations. In the case without turbulent broadening, the circumstellar gas only slightly reduces the measured log\,$R_{\text{HK}}'$. However, when turbulent broadening is included, the circumstellar gas can greatly reduce the log\,$R_{\text{HK}}'$ further separating the two populations.}

Without turbulent broadening, the reduction in the measured log\,$R_{\text{HK}}'$ is minor because, although the circumstellar gas is optically thick at line centre, the low thermal velocity of calcium atoms means the gas becomes optically thin rapidly away from line centre. The torus covers a large portion of the stellar disk, but due to Keplerian motion, different regions of the torus have different line-of-sight velocities relative to the observer. These velocity shifts mean that gas in one part of the torus may absorb strongly at a particular wavelength, while gas in another region does not absorb strongly at that wavelength. As a result, the optically thick area at any single wavelength is small, leading to only minimal total absorption in the line profile. Aside from turbulent broadening, a higher column density of gas (beyond the EUV ionization limit adopted in this work), would also lead to stronger absorption.  

\section{The effect of circumstellar gas on planetary transits}

Since a circumstellar torus alters the observed stellar spectrum, it may also modify the observed transit of the planet, whose mass loss produces the circumstellar gas. This impact is limited to the wavelengths that are strongly attenuated by the circumstellar gas, namely, electronic transitions from atoms and ions, which typically probe the planet's thermosphere or exosphere. 

To investigate this, we performed synthetic transit simulations of a planet undergoing atmospheric escape in the He\,\textsc{i} 10830~\AA~line, incorporating the opacity from a circumstellar torus. In our approach, we separate the long-lived circumstellar gas, modeled as described in Section~\ref{sec:model}, from the planetary outflow confined within the Hill sphere. The outflow inside the Hill sphere is simulated using \textsc{p-winds} \citep{DosSantos2022}, which treats the outflow as an isothermal Parker wind and calculates the steady-state distribution of helium in the outflow. Both the circumstellar torus and the planet are assumed to have an impact parameter of zero. The planetary and stellar parameters used in the simulations are listed in Table~\ref{tab:escaping_system_parameters}. We compute the optical depth of the outflow as a function of position on the stellar disk, following the method described in Section~\ref{sec:synthetic_transit}.

\begin{table}
  \caption{The system parameters used to perform synthetic He\,\textsc{i} 10830~\AA~transit observations for a planet whose mass loss generates a circumstellar torus. The spectrum used was that of the K-star HD 97658 from the MUSCLES treasury survey \citep{France2016, Loyd2016}.}
  \begin{tabular}{p{4cm}p{2cm}}
   Stellar Mass (M$_\odot$) & 0.77\\
   Stellar Radius (R$_{\odot}$) & 0.72\\ 
   Planetary Mass (M$_\text{J}$) & 0.7\\
   Planetary Radius (R$_\text{J}$) & 1.4\\
   Planetary Semimajoraxis (AU) & 0.023\\
   Planetary Mass-Loss Rate (g s$^{-1}$) & $3 \times 10^{11}$\\
   Outflow Temperature (K) & 8000\\
   Torus Optical Depth to EUV & (0, 0.2, 2)\\
   Torus Inclination ($^{\circ}$) & 90\\
   Torus Composition & Solar 
   \end{tabular}
   \label{tab:escaping_system_parameters}
\end{table}

\begin{figure*}
\centering
\includegraphics[width=\textwidth]{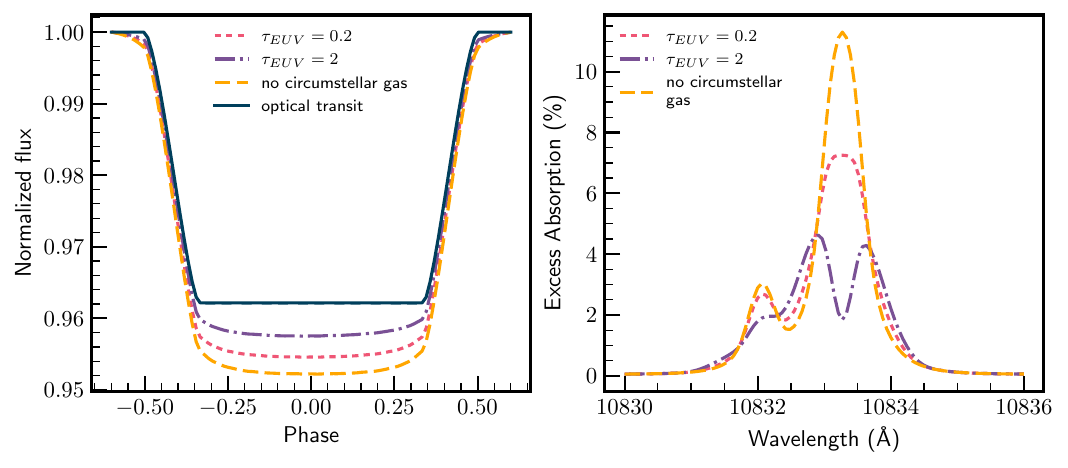}
\caption[Transit]{The synthetic He\,\textsc{i} 10830~\AA~transit for a planet with an escaping H/He atmosphere orbiting a star surrounded by a circumstellar torus. The left panel displays the photometric transit for varying optical depths of the circumstellar gas, while the right panel presents the corresponding mid-transit spectra. In both cases, increasing the optical depth of the circumstellar torus reduces the observed depth of the planetary transit.}
\label{fig:transit_spectrum}
\end{figure*}

Figure~\ref{fig:transit_spectrum} shows both the photometric transit of the planet and the transmission spectrum at mid-transit for different EUV optical depths of the circumstellar torus, which serve as a proxy for its density. The photometric transit is obtained by integrated the spectrum in a 6 \AA~bin centered at 10833.27 \AA, the weighted average of the two strongest transitions of the He\,\textsc{i} $2^3$s $\rightarrow 2^3$p triplet. As the circumstellar gas becomes denser, the transit signal becomes increasingly muted, although its overall shape remains largely unchanged. In extreme cases, the gas could suppress the signal enough to render the transit undetectable. 

The transmission spectrum at mid-transit shows a similar trend. As the circumstellar gas becomes denser, the depth of the transit at line centre decreases. However, the shape of the absorption feature is also affected. At intermediate densities, the relative strength of the two peaks of the triplet decreases. For optically thick circumstellar gas (to 10830 \AA~radiation), the transit at line centre becomes completely obscured, and instead, absorption signals appear at unexpected locations away from the line centre.

\section{Discussion}

Previous studies have established a link between the log\,$R'_{\text{HK}}$ of stars and the presence of close-in giant planets \citep{Hartman2010, Figueira2014, Staab2017, DohertyThesis}. One proposed explanation for this link is that gas evaporated from a close-in planet forms a circumstellar torus, which absorbs in the cores of the Ca\,\textsc{ii} H and K lines. This absorption reduces the observed flux at all orbital phases, thereby lowering the measured 
log\,$R'_{\text{HK}}$ \citep[e.g.,][]{Haswell2012, Lanza2014, Staab2017, DohertyThesis}.

In this work, we model circumstellar gas produced by mass loss from close-in planets, {\rc assuming that the density of the gas is such that it is marginally optically thick to EUV radiation. We find that the circumstellar gas significantly deepens the observed He\,\textsc{i} 10830 \AA~line. To produce a substantial reduction in the observed log\,$R'_{\text{HK}}$, there needs to either be some degree of extra broadening of the Ca\,\textsc{ii} H \& K absorption, whether due to turbulent velocities or another mechanism, or the column density of circumstellar gas needs to be substantially larger than modelled. This is necessary for our results to align with the interpretation proposed by \citet{Haswell2012, Staab2017}, namely that Ca\,\textsc{ii} absorption in circumstellar gas is responsible for the anomalously low log\,$R'_{\text{HK}}$ values in some stars. Nevertheless, we have shown that complementary observations of the stellar He\,\textsc{i} 10830 \AA~and Ca\,\textsc{ii} H \& K lines are an effective way to identify circumstellar tori produced from the mass loss of close-in planets.} 

We therefore propose a survey of the He\,\textsc{i} 10830 \AA~line in stars with close-in planets (with observations taken out of transit), complemented by Ca\,\textsc{ii} H \& K measurements to test the hypothesis that mass loss from close-in planets form circumstellar tori of gas. The sample should include both transiting and non-transiting planets. {\rc Some of these non-transiting systems can serve as a control sample, if they are inclined such that any circumstellar torus would not be aligned with our line of sight.}

Since the He\,\textsc{i} 10830~\AA\ line is a common tracer of atmospheric escape, there are already some observations of both He and Ca lines for stars hosting close-in planets (obtained while the planet is out of transit). Many of these planets are too distant from their host stars to undergo the extreme mass loss required to produce a circumstellar torus. However, a few systems, such as WASP-12, HAT-P-32, and HAT-P-67, have both a stellar He\,\textsc{i} 10830 \AA~EW and a log\,$R'_{\text{HK}}$ measurement (out of the supposed transit), and a close-in giant planet that been observed undergoing, extreme atmospheric escape \citep[][]{Haswell2012, Czesla2022, Zhang2023, Gully-Santiago2023}. These systems can be placed within the He\,\textsc{i} 10830 \AA~EW-log\,$R'_{\text{HK}}$ parameter space. 

WASP-12 has a He\,\textsc{i} 10830 EW of $\sim 100$ m\AA\ and and a log\,$R'_{\text{HK}}$ of $\sim -5.5$ \citep{Czesla2024}. {\rc These values are inconsistent with that of a typical field star (see Figure~\ref{fig:he_ca_correlation}, \citealt{Smith2016}), but, according to our model, are consistent with WASP-12 (a late-F star) hosting a circumstellar torus with significant turbulent velocities.}  

HAT-P-32 has an He\,\textsc{i} 10830 EW of $\sim$ 250 m\AA\ \citep{Zhang2023} and a log\,$R'_{\text{HK}}$ of roughly $-4.6$ \citep{Claudi2024}, while HAT-P-67 shows an He\,\textsc{i} EW of $\sim$ 200 m\AA\ and log\,$R'_{\text{HK}}$ of approximately $-4.7$ \citep{Sicilia2024}. These measurements are consistent with those of typical field stars \citep{Smith2016}, hinting that the observed stars are unobscured. We also note that recent hydrodynamics simulations are able to explain the extended helium outflow observed in the systems \citep[][]{Nail2025}. In-transit, both HAT-P-32 and HAT-P-67 exhibit EWs $\sim 350$ m\AA. If this value was taken as the true stellar helium line, both these stars would appear anomalous in He\,\textsc{i} 10830 \AA~EW-log\,$R'_{\text{HK}}$ space, and more in line with expectations for systems hosting circumstellar tori (both are F-type stars).

As discussed in Section~\ref{sec:model}, the formation of a circumstellar torus depends not only on the planet's mass-loss rate, but also on stellar wind strength, magnetic field and radiation pressure. In addition, the long-term stability of such a structure depends on the gas viscosity and its dynamical interactions with the planet. Therefore, the existence or absence of these tori can place meaningful constraints on key system parameters. In Section~\ref{sec:model}, we found that the formation of a circumstellar tori requires the planetary mass-loss rate to exceed that of the stellar wind. Typical mass-loss rates for close-in planets range from $10^{9} -10^{13}$ g s$^{-1}$ \citep[e.g.,][]{Murray-Clay2009, GarciaMuniz2007}, which overlaps with the range of estimated stellar wind mass-loss rates \citep{Wood2005b}. Thus, the search for circumstellar tori serves as a powerful probe of stellar wind properties, both for individual stars and across the broader stellar population. 

A caveat to the preceding discussion is that we must ensure: (1) our conditions for the formation of circumstellar gas are correct, and (2) the circumstellar gas indeed produces the observational signatures we predict. Fully verifying this requires three-dimensional hydrodynamic simulations that incorporate the many relevant physical processes. In the following discussion, we outline the limitations of our current model and directions for future work.

\subsection{Limitations and Future Work}

\subsubsection{Variations in the structure of the circumstellar gas}

Our simplified model does not self-consistently calculate the density and velocity structure of the circumstellar gas. Because there are no long-term, global hydrodynamic simulations of circumstellar torus formation and evolution, the true structure of the gas remains uncertain, and deviations from our assumed configuration are likely. For example, high viscosity could cause the gas to spread into a more extended disk-like structure, while gravitational interactions with the planet might carve gaps. The stellar wind, though perhaps insufficient to clear the gas, could cause it to slowly spiral outward, and stellar magnetic fields could torque the gas, altering its orientation. 

The key question is whether deviations from our simplified model could significantly affect the expected observational signatures of the circumstellar gas. Assuming the gas is present, we argue that the strong He\,\textsc{i} 10830~\AA\ absorption remains a robust prediction. Firstly, even when we reduce the EUV optical depth of the circumstellar gas by an order of magnitude in our models, we still find substantial He\,\textsc{i} 10830~\AA\ absorption. Therefore, there are only two plausible ways to significantly reduce the optical depth at 10830~\AA~while retaining the presence of circumstellar gas: 1) the gas is depleted in helium; or 2) the population of helium in the He\,\textsc{i} $2^3$s is extremely low. Helium is expected to be abundant in the atmospheres of gas-rich planets and is a light element, so it would be carried along in the vigorous outflows that can generate a circumstellar torus. We note that this is not necessarily the case for less vigorous escape from smaller planets, where differentiation is predicted \citep[e.g.,][]{Malsky2023, Schulik2025}. Therefore, it is unlikely that the circumstellar gas would be significantly depleted in helium. 

We also do not think that the fraction of helium in the $2^3$s state could be depleted compared to our estimate. In steady state, the fraction of helium in the $2^3$s state can be approximated as \citep{Schulik2025}:
\begin{align}
X_{\text{He\,\textsc{I} }2^3\text{s}} = \frac{\alpha(\text{He\,}\textsc{i } n^3\text{l})}{q_{31} + \frac{1}{\Gamma(\text{He\,\textsc{i} }2^3\text{s})n_e}}X_{\text{He\,II}}
\end{align}
where $\alpha_{\text{A}}(\text{He,}\textsc{i }n^3\ell)$ is the Case A recombination coefficient into the triplet state, $q_{31}$ is the coefficient for collisional depopulation of the triplet state by electrons (both given in Table~\ref{tab:Helium State Processes}), and $\Gamma(\text{He,\textsc{i} }2^3\text{s})$ is the triplet photoionization rate. At $10^4$ K, for electron number densities of $\sim 10^{6}$–$10^{8} \text{ cm}^{-3}$ and typical triplet photoionization rates of $10^{-2}$–$10^{0} \text{ s}^{-1}$, the fractional population of helium atoms in the He\,\textsc{i} $2^3$s state remains $\gtrsim 10^{-7}$. Significant depletion of this level occurs only at very low electron densities, where recombination into the $2^3$s state becomes inefficient. The only other way to strongly suppress the $2^3$s population is to make the gas predominantly neutral, which is not physically realistic in our parameter space.

We have less confidence in the results regarding how the circumstellar torus affects the stellar Ca\,\textsc{ii} H \& K line cores, and thus the measured log\,$R'_{\text{HK}}$. Calcium is a relatively heavy element (atomic mass $\approx$ 40), so it may not escape the planet efficiently and could be absent or severely depleted in the circumstellar torus. {\rc In this case, the stellar Ca\,\textsc{ii} H \& K line cores would be unaffected by the circumstellar gas. Alternatively, it is possible that the attenuation of these line cores is stronger than expected. As shown in Figure~\ref{fig:circumstellar_comparison}, large turbulent velocities broaden the Ca\,\textsc{ii} H \& K absorption profile, leading to a stronger attenuation of the stellar line.}

\subsubsection{Biases}

We have assumed that the intrinsic relationships between log\,$R'_{\text{HK}}$, He\,\textsc{i} 10830 \AA~EW, and other stellar activity metrics are the same for stars in the \citet{Smith2016} sample (with $B-V > 0.47$), which we take to represent field stars, and for stars hosting close-in planets that may possess circumstellar tori. The He\,\textsc{i} 10830~\AA\ equivalent width measurements in this sample are predominantly drawn from \citet{Zarro1986} and the log\,$R'_{\text{HK}}$ values were compiled from various sources but were homogenized following the methodology described in \citet{Smith2011}. Stars with both He\,\textsc{i} 10830~\AA EW and log\,$R'_{\text{HK}}$ measurements are relatively sparse (covering $\sim$50 stars), therefore will not cover the full stellar parameter space. 

This is problematic because we do not expect stars hosting close-in planets to have the same properties as field stars. One well-known difference is that stars hosting giant planets tend to be, on average, more metal-rich than typical field stars \citep[e.g.,][]{Fischer2005, Petigura2018}. It is established that log\,$R'_{\text{HK}}$ is negatively correlated with stellar metallicity \citep{Rocha-Pinto1998}. Assuming the He\,\textsc{i} 10830~\AA\ equivalent width is not dependent on metallicity for a given level of stellar activity, high-metallicity stars would exhibit lower log\,$R'_{\text{HK}}$ values compared to lower-metallicity stars with the same He\,\textsc{i} 10830~\AA\ equivalent width. This would appear as a deficit in the mean log\,$R'_{\text{HK}}$ of stars hosting close-in planets compared to field stars.  

For the reasons outlined above, there may be inherent differences between the intrinsic properties of the \citet{Smith2016} sample and stars hosting close-in planets. More work is needed to account for these potential biases. However, we do not think this affects the ability to test for the presence of circumstellar gas, as the He\,\textsc{i} 10830~\AA\ absorption from such gas is predicted to be very strong, significantly stronger than any variation we would expect from population differences or selection effects.

\subsection{Application of model to other spectral lines}

\begin{figure*}
\centering
\includegraphics[width=\textwidth]{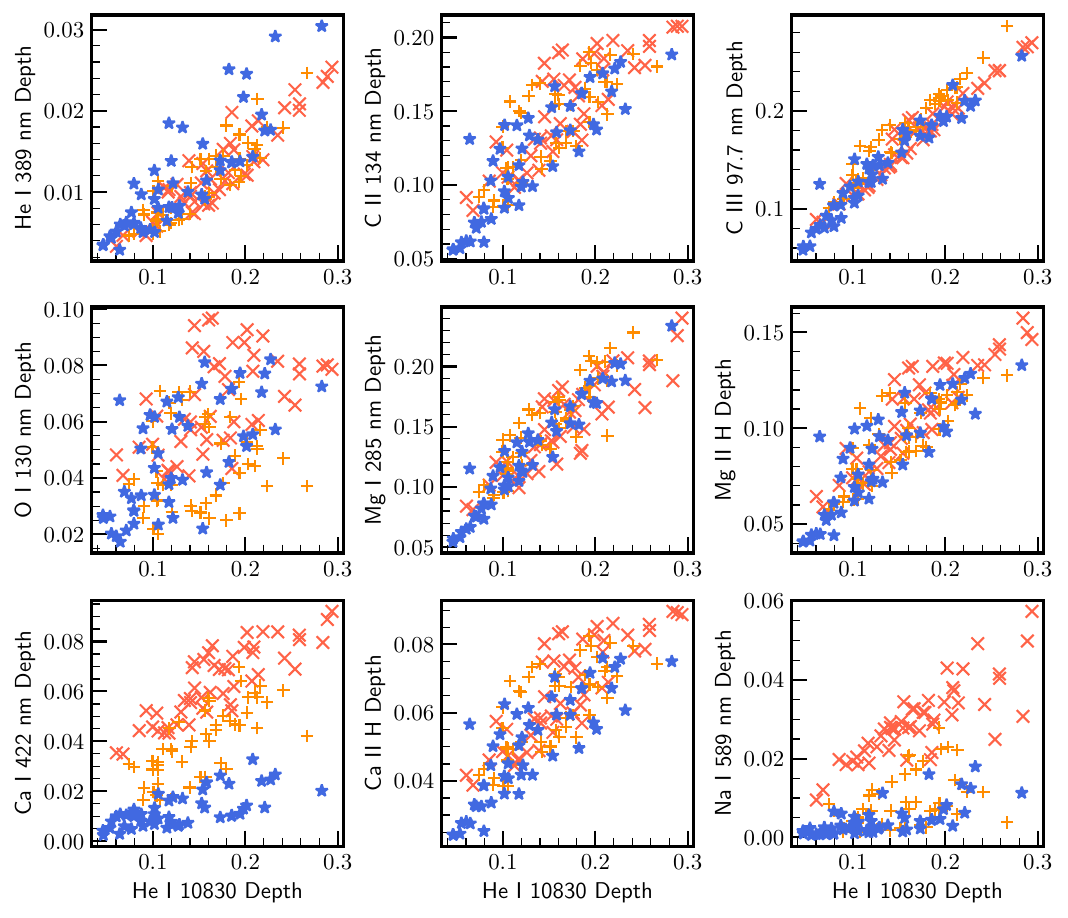}
\caption{{\rc The fractional absorption of starlight by the circumstellar torus is shown for various spectral lines for the same sample of stars hosting circumstellar tori as in Figure~\ref{fig:circumstellar_comparison}). Blue stars, yellow pluses and red crosses are F,G and K stars respectively. The absorption is calculated over a 100 km s$^{-1}$ interval around the line center, and the gas is assumed to have no turbulent velocities. Some lines show significant absorption, reaching up to 20\% of the stellar flux.}}
\label{fig:all_lines}
\end{figure*}

In this work, we have focused on the He\,\textsc{i} 10830~\AA~and Ca\,\textsc{ii} H \& K lines, however other spectral lines also offer avenues to detect and characterize the circumstellar gas. Most resonant absorption lines from hot atomic or ionized gas lie in the UV, requiring space-based observations. As a result, available data regarding the properties of stars at these wavelengths is limited. Despite this, there are some notable empirical correlations between chromospheric, transition-region, and coronal emissions for F–K stars \citep[e.g.,][]{Ayres1981}. In principle, a similar methodology to the one used for the  He\,\textsc{i} 10830~\AA~and Ca\,\textsc{ii} H \& K correlation could be applied to assess which wavelengths to probe for evidence of circumstellar gas. We leave such studies to future work. However, to guide which spectral transitions may be most promising to investigate, in Figure~\ref{fig:all_lines} we show the simulated occultation depth of the circumstellar gas for different transitions for the same sample systems as in Section~\ref{sec:rings}. In this calculation, the stellar line is assumed to be flat, and the photometric depth is computed over a 100 km s$^{-1}$ interval. {\rc The gas turbulent velocity is set equal to zero.} 

\section{Conclusions}

In this work, we developed a model of circumstellar gas tori generated by the mass loss from close-in planets. Using this model, we investigated how the presence of such tori  modifies the stellar He\,\textsc{i} 10830~\AA\ and Ca\,\textsc{ii} H \& K lines. We find that stars hosting circumstellar tori can be distinguished from those without, primarily due to unusually strong He\,\textsc{i} 10830~\AA\ absorption relative to their chromospheric Ca\,\textsc{ii} H \& K emission. This enhanced absorption arises from dense circumstellar gas produced by planetary mass loss. Consequently, observational surveys targeting the He\,\textsc{i} 10830~\AA\ line in stars with close-in planets, when combined with new or existing measurements of Ca\,\textsc{ii} H \& K emission, offer a powerful means of testing for the presence of circumstellar tori. In doing so, can provide valuable insight into the evolution and mass loss processes of close-in planets.

\section*{Acknowledgements}

The authors acknowledge support from NASA XRP grants 80NSSC23K0282 and 80NSSC25K7153. ES has received funding from the European Research Council (ERC) under the European Union’s Horizon 2020 research and innovation programme (Grant agreement No. 853022, PEVAP). RMC acknowledges support from HST-GO-17157. This work benefited from the 2025 Exoplanet Summer Program in the Other Worlds Laboratory (OWL) at the University of California, Santa Cruz, a program funded by the Heising-Simons Foundation and NASA. We thank James Owen, Anna Penzlin, Subuhanjoy Mohanty and Carole Haswell for helpful comments that improved the quality of the manuscript. 

\section*{Data Availability}

The code used to generate these results is available at \href{https://github.com/eschreyer/torus}{github}. 



\bibliographystyle{mnras}
\bibliography{Bibliography} 




If you want to present additional material which would interrupt the flow of the main paper,
it can be placed in an Appendix which appears after the list of references.


\bsp	
\label{lastpage}
\end{document}